\documentclass{acm_proc_article-sp}
\usepackage{epstopdf}
\usepackage{color}
\usepackage{stfloats}
\usepackage{hyperref}
\usepackage{pifont}% http://ctan.org/pkg/pifont
\newcommand{\cmark}{\ding{51}}%
\newcommand{\xmark}{\ding{55}}%
\makeatletter
\def\@copyrightspace{\relax}
\makeatother
\begin{document}

\title{A case study in formal verification of a Java program}

\numberofauthors{2}
\author{
\alignauthor
Dmitry Brizhinev\\
       \affaddr{Research School of Computer Science}\\
       \affaddr{Australian National University}\\
       \affaddr{Canberra, Australia}\\
       \email{\normalsize dmitry.brizhinev@anu.edu.au}
\alignauthor
Rajeev Gor\'{e}\\
       \affaddr{Research School of Computer Science}\\
       \affaddr{Australian National University}\\
       \affaddr{Canberra, Australia}\\
       \email{\normalsize rajeev.gore@anu.edu.au}
}

\maketitle
\begin{abstract}
We describe a successful attempt to formally verify a simple genetic
algorithm written in Java. To this end, we compare several formal
verification tools designed for Java, and select Krakatoa as the most
appropriate for the task. Based on our experience, we present several
suggestions for making the tools more user friendly, which we hope
will lead to wider adoption of formal methods. In particular, we
discuss at length how useful it would be for provers to perform some
form of \emph{abduction}, that is, for them to guess which extra
assumptions they need to prove a statement. It is our opinion that
progress in this area would produce the largest improvement in the
usability of formal verification tools.
\end{abstract}

\section{Introduction}

Formal verification is a family of techniques for constructing proofs of the correctness of software. A program with a formal proof of correctness is much less likely to contain bugs than one that has merely been tested. Unfortunately, the process is still prohibitively difficult with current tools, and formal verification is thus confined to systems with a very high cost of failure, such as CPU design \cite{intel} and aircraft control systems \cite{aircraft}. However, we hope that better tools could add formal verification to the average programmer's code-checking toolkit, alongside testing and code review, thus improving code quality across the board.

Additionally, formal methods are easier to apply and have been better developed for functional programming languages such as Haskell than for the more commonly used imperative languages such as C and Java. As an example, the seL4 project to verify an operating system kernel \cite{sel4} used Haskell as an intermediate representation in their proof process, rather than verifying properties of the C program directly. As we are interested in encouraging wider adoption of formal methods, we investigated the current state of the art for verification of Java programs, a language currently seen as first or second by popularity \cite{topten, tiobe, redmonk}. To this end, we compared several verification tools for Java, and then verified a simple program using the tool we thought the most appropriate, Krakatoa \cite{krakatoa}.

In this paper, we provide a comparison of the tools we investigated, an overview of our experience with Krakatoa, and several suggestions for making formal tools more accessible.
\newpage

\subsection{JML}

JML (Java Modelling Language) \cite{jml} is a language for annotating Java programs with specifications. It supports function pre- and post-conditions, loop invariants, class invariants, and assertions at arbitrary positions in the code. JML annotations are specially-formatted comments. Its syntax is very similar to Java code, making it very easy for a Java programmer to learn and apply, and it supports some advanced operations such as \verb$\forall$ and \verb$\sum$.

Java verification tools universally assume specifications written in JML, although they vary in their level of support for particular JML constructs.

\subsection{Models}

Reasoning about the behaviour of a program requires encoding the
semantics of the programming language into logic formulae
\cite{modelchecking}. This encoding is called a model. If the model
and the properties of the program we wish to prove are expressed as an
SMT problem, we can send them to an off-the-shelf solver that will attempt to prove the assertion or find a counterexample.

\subsection{SMT}

Traditional SAT solvers reason about formulas expressed in pure propositional logic. However, statements we wish to prove about programs often involve additional constructs with associated theories, such as the theory of integers, or the theory of lists. We can thus consider formal verification as an SMT (satisfiability modulo theories) problem \cite{smt}. While it may be possible to express a theory as a set of propositional axioms over the objects of interest, it turns out to be more efficient to extend SAT solvers with special-purpose routines for reasoning about these theories. Many such SMT solvers are available, including CVC4 \cite{cvc4}, Simplify \cite{simplify}, Yices \cite{yices}, and Z3 \cite{Z3}. 

\subsection{Types of tool}

Given a Java program annotated with JML, there are several possible approaches for checking its correctness.

\subsubsection{Runtime assertion checking}

JML pre-conditions, post-conditions, class invariants and assertions can be checked at runtime. This is not really formal verification and will not be further considered here.

\newpage

\subsubsection{Bounded model checking}

Bounded model checking verifies the correctness of the program
assuming some fixed bound on the number of times a loop body can be
executed and a bound on the size of any arrays. Under these
assumptions, either a proof of correctness or a counterexample can
usually be generated automatically, even for complicated
specifications. The result is equivalent to testing all possible
inputs up to some size limit. It is hoped that most bugs in the
program will manifest themselves on small examples. This is the
easiest way to verify a program, but cannot guarantee correctness for
inputs (or execution paths) that exceed the bounds.
\subsubsection{Automated SMT solving}

Some tools encode the program and the JML assertions as an SMT problem and send it to an off-the-shelf SMT solver. For this kind of proof to succeed, loops need to be annotated with appropriate invariants. Even then, between proof and counterexample lies a large space of inputs on which the SMT solvers will time out or report failure. A verification tool based on automated SMT solving (only) will then also report failure. The only way to respond is to change the annotations to something that might be easier to prove, or to abandon the proof attempt. In our experience, it is usually possible (but difficult) to find annotations that lead to success. The automated SMT solving approach is sometimes referred to as ESC (Extended Static Checking).

\subsubsection{Interactive SMT solving}

Some tools can also encode the SMT problem in a format suitable for
sending to an interactive proof assistant such as Isabelle/HOL
\cite{isabelle} or Coq \cite{coq}. KeY\cite{key} comes with its own interactive proof environment. These tools can still send proof obligations to an automated solver, but when the attempt fails they allow the user to try completing the proof interactively. We thus refer to them as Automated/Interactive SMT tools.

\subsubsection{Embedding in a theorem prover}

It is also possible to port the entire program to the native language
of Isabelle/HOL or Coq, and prove statements about this
version. Several Java programs have been verified in this way
\cite{acl2,jml4}. This gives maximum proving power, but potentially
sacrifices the ease of the automated tools. There is also no guarantee
that the Isabelle/HOL version of the program accurately captures all
the semantics of the source language (e.g. integer overflows). The
approach of seL4 \cite{sel4} is similar to this -- a separate proof of
the behaviour of a Haskell program, and of the correctness of a C
program with respect to this Haskell specification.

\section{Comparison}
We first present a side-by-side comparison of several tools, and then
more detailed comments on our experience with each tool. In some
cases, we use simple subjective judgements rather than something precise like ``man-hours'' because different users are likely to work at different rates.

\subsection{Brief comparison}

Tables \ref{overview}, \ref{support} and \ref{use} give an overview of
the tools we compared. The columns are described below, with ``?''
meaning that we could not download the tool or we could not get it to run.
\newpage
\begin{description}
\item [Available] Was the tool available for download?
\item [Maintained] Are the tool's developers still involved, or has it been abandoned?
\item [Type] The type of tool: BMC (Bounded Model Checker), A-SMT (Automated SMT solver), A/I-SMT (Interactive SMT solver).
\item [Java version] An indication of the Java versions supported by the tool.
\item [FP] Can the tool reason about floating-point variables?
\item [\textbackslash forall] Does the tool understand the JML \verb$\forall$ annotation? It is useful for reasoning about arrays.
\item [\textbackslash sum] Does the tool understand the JML \verb$\sum$ annotation? It is useful for proving termination of complex loops.
\item [Setup] Subjective judgement of how difficult we found it to install the tool and get it to work.
\item [Use] Subjective judgement of how difficult we found it to use the tool on several toy programs.
\end{description}

\begin{table}[t!]
\caption{Overview of Java verification tools surveyed}
\begin{tabular}{lllll}
 Tool &  Available & Maintained & Type & Ref \\
 \hline\\
 LOOP &  \xmark & \xmark & ? & \cite{loop} \\
 JACK &  \xmark & \xmark & ? & \cite{jack} \\
 TACO & \cmark & ? & BMC & \cite{taco} \\
 Sireum/Kiasan & \cmark & ? & BMC & \cite{kiasan} \\
 ESC/Java2 & \cmark & \xmark & A-SMT & \cite{escjava2} \\
 OpenJML & \cmark & \cmark & A-SMT & \cite{openjml} \\
 KeY & \cmark & \cmark & A/I-SMT & \cite{key} \\
 Krakatoa & \cmark & \xmark & A/I-SMT & \cite{krakatoa}
\end{tabular}
\label{overview}
\end{table}

\begin{table}[t!]
\caption{Capabilities}
\begin{tabular}{lllll}
 Tool &  Java version & FP & \verb$\forall$ & \verb$\sum$ \\
 \hline\\
 LOOP &  ? & ? & ? & ? \\
 JACK &  ? & ? & ? & ? \\
 TACO & $=$ 6 & \xmark & \cmark & \cmark \\
 Sireum/Kiasan & $=$ 6 & \xmark & \xmark & \xmark \\
 ESC/Java2 & $=$ 4 & ? & ? & ? \\
 OpenJML & $=$ 7 & \xmark & \cmark & \xmark \\
 KeY & $\geq$ 6 & \xmark & \cmark & \cmark \\
 Krakatoa & Any & \cmark & \cmark & \xmark 
\end{tabular}
\label{support}
\end{table}

\begin{table}[t!]
\caption{Ease of use}
\begin{tabular}{lll}
 Tool & Setup & Use \\
 \hline\\
 LOOP & ? & ? \\
 JACK & ? & ? \\
 TACO & Difficult & Very Easy \\
 Sireum/Kiasan & Difficult & Easy \\
 ESC/Java2 & Very Difficult & ? \\
 OpenJML & Difficult & Average \\
 KeY & Average & Very Difficult \\
 Krakatoa & Very Difficult & Easy
\end{tabular}
\label{use}
\end{table}

\clearpage

\subsection{Detailed observations}

\subsubsection{LOOP and JACK}

LOOP and JACK seem to have been popular and very promising in the early 2000s \cite{microsoft}, but LOOP no longer has a website and JACK's download links no longer work:
\newline
\url{http://www-sop.inria.fr/everest/soft/Jack/jack.html}

\subsubsection{TACO}
This tool takes a little effort to set up, but it illustrates the benefits of bounded model checking. It easily proves (bounded versions of) statements that the more powerful tools struggle with. It does not support floating point arithmetic and knows it, producing an error message:
\newline
\url{http://www.dc.uba.ar/inv/grupos/rfm_folder/TACO}

\subsubsection{Sireum/Kiasan}
Sireum/Kiasan is another bounded model checker that we felt was inferior to TACO. It claims to support floating point, but in fact does not (it proves the statement \texttt{0.0 == 1.0}). It also does not support the \verb$\forall$ annotation, which is vital for reasoning about arrays:
\newline
\url{https://code.google.com/archive/p/sireum/}

\subsubsection{ESC/Java2}
This tool seems to have been popular in its time, but has been
superseded by OpenJML. It was difficult to install and we were unable
to get it to run. Supposedly it can send proof obligations to
Coq, which might elevate it to A/I-SMT status (OpenJML cannot do this
at the present time): \newline
\url{http://kindsoftware.com/products/opensource/ESCJava2/}

\subsubsection{OpenJML}
The state of the art automated SMT tool. We found it difficult to get working, but eventually succeeded. Unfortunately, it was not very useful. When attempting a proof, OpenJML sends an SMT problem to a solver. If the solver fails, it gives up. The only action available afterwards is to edit loop invariants, in the hope that something will make the prover succeed. For proving simple properties, however, OpenJML is quite easy to use. It does not support floating point and produces an error message:
\newline
\url{http://www.openjml.org/}

\subsubsection{KeY}
KeY is the most prominent interactive tool for Java verification. It has its own interactive proof environment, and supports sending proof obligations to SMT solvers. However, we found it extremely difficult to accomplish anything in the tool. The documentation was several years (and several major versions) out of date. We were not able to actually launch an SMT solver from within KeY. We also could not introduce a lemma in the interactive environment. KeY had patchy support for some JML features -- in particular, we did not know how to specify that some element of an array is not modified in a loop body. Finally, KeY has a floating point option of questionable utility, unable as it is to prove that \texttt{0.0 == 0.0}:
\newline
\url{http://www.key-project.org/}

\newpage
\subsubsection{Krakatoa}
Krakatoa is the most promising tool of the ones we have tried. It is built on \emph{Why3} \cite{why3}, a programming language and toolchain designed for verification. Krakatoa leverages the existing toolchain to convert Java programs and JML annotations into SMT problems, or Coq or Isabelle proof goals. Krakatoa is the only tool we found that can reason correctly about floating point variables. It can treat them as mathematical real numbers, or use a third-party Coq library that describes the precise behaviour of IEEE floating point.

Within the tool, the user can split proofs into separate
obligations. The tool can display which code block each obligation
refers to, and can send them to SMT solvers separately. This produces
several benefits. Firstly, if one proof attempt fails and another
succeeds, the user knows which aspect of the code to focus
on. Secondly, users might decide that they are satisfied with
leaving some aspect unproven. The tool also provides separate goals
for e.g. invariant preservation and termination, and separate goals
for null pointer checks, array bounds checks, and integer
overflows. Thus, a user can choose to ignore these error conditions if
they wish. Many of the other tools allow us to turn off integer
overflow checking (for simplicity), but only before
beginning the proof.

Unfortunately, Krakatoa does not support the JML \verb$\sum$ and \verb$\max$ statements. These are very helpful for proving termination of a loop (as part of the \emph{variant} that decreases with each loop iteration). However, Krakatoa does technically allow the inductive definition of complex logical functions (which are converted directly into Coq functions), so the sum could be defined in this way.

Finally, we found it extraordinarily difficult to actually install Krakatoa. Although the tool is currently "frozen" while the team works on Why3, it gives the appearance of being well maintained. However, the install scripts seem to be broken, and the tool failed to automatically install its dependencies and Coq libraries. It took much effort to run the tool, but the result was pleasing. We thus decided to use Krakatoa for our verification task:
\newline
\url{http://krakatoa.lri.fr/}

\section{Verification}

The Java program we verified is available at:\newline\url{https://github.com/dmitry-brizhinev/formal-verification}.\newline
It is a
simple genetic algorithm that investigates different deterrence
strategies in international conflict. This computational model may be
the subject of a future paper.

\subsection{Computational modelling}

Computational modelling in the sciences involves running a simulation of some kind of system and observing the results. If the assumptions that went into the system hold in the real world, it is hoped that any insights gained from observing it will carry over to an understanding of the real-world phenomena of interest.

The functional correctness of such models is an issue not often addressed. If a model with assumptions X produces output Y, we want to be confident that it is the dynamics of the model that led to Y, not a bug. However, since the \emph{desired} output of the model is unknown, such bugs are often not obvious, and it is not necessarily possible to construct test cases. Indeed, the model may produce output that the experimenter expects, and thus incorrectly confirm their preconceptions, due to a bug. Formal methods are a natural solution to this problem.

\subsection{Specification}

The first step in formally verifying a piece of software is to specify its behaviour in an appropriate form. The verification process then provides a proof that the program's real behaviour adheres to the specification.

Unfortunately, the program in question relies heavily on random
numbers. Thus, an ideal specification would need to express statements
such as \emph{this function correctly calculates the expected value of
  this random variable}. As it stands, the Java Modelling Language
provides no means to reason about random numbers, so the best specification of the function's behaviour is the code itself. When no separate specification is possible, formal verification adds no value. Instead, the verification effort focused on proving that no exceptions could be raised during execution, and on a few basic assertions (such as: that the expected value of an event with two possible outcomes is somewhere between the outcomes).

Each function was annotated with pre- and post-conditions that ensured all objects and arrays used by the code were valid at all times. There were also a few assertions ensuring that the output values of functions were reasonable.

Full verification of this kind requires the use of appropriate loop invariants. These were added as they became necessary during the course of the proof.

\subsection{Workflow}

The Krakatoa tool takes a Java source file annotated with JML assertions and produces a list of Why3 obligations. The Why3 tool then provides an interface for proving these obligations. It can perform some basic manipulations (such as splitting an obligation into parts), send obligations to automated provers (such as the SMT solvers CVC4, Z3, Simplify, and Yices), or create proof scripts for interactive provers (such as Coq or Isabelle). Why3 also makes an effort to display which line of Java code each obligation refers to, which assertion (or safety property) it is checking, and which assertions it includes as assumptions.

Verifying code using Krakatoa is a trial-and-error process. The initial JML annotations are unlikely to be sufficient. Having observed the proof goals generated by the tool, and which of them the automated provers cannot solve, one returns to the code and adds annotations as necessary. There is a lot of boilerplate in the final product. For example, many functions come with post-conditions guaranteeing that the function does not deallocate the objects it works with. Sometimes a proof goal is out of reach of the automated provers but can be discharged interactively in Coq. Most of the time, however, an attempt at a proof in Coq simply illustrates which assumptions are missing. Once the relevant assertions are added, the automated prover succeeds.

\newpage
\subsection{Result}

The correctness and safety of the model was mostly proven. We did not find any errors. There are three gaps (unfulfilled obligations) in the proof.

Firstly, although Krakatoa correctly adds the values of \texttt{static final int} constants as assumptions, it ignores the values of \texttt{static final double} ones. Thus, the value of one constant had to be added as a precondition to the \texttt{main} method for the proof to succeed.

Secondly, Krakatoa provides very little to work with upon entering an
object constructor. It is impossible to prove that the new
object being constructed is not equal to any other object already in
existence. This thus remains an unproven assertion in the program. It
is not true in general (if the object has a superclass, the superclass
constructor may have already assigned the object to something else in
the environment), Krakatoa simply does not provide any assumptions
that would allow one to prove the statement even in specific cases
where it is true. Notably, the software includes a few test cases for
situations akin to this, but the test cases are very weak -- Krakatoa
would fail to prove almost anything more complicated than those
specific cases.

Finally, Krakatoa is unable to prove that \texttt{System.out != null},
which makes it impossible to verify the safety of a \texttt{print}
statement.

All other proof obligations generated by Krakatoa have been
proven. 506 out of 512 were completed automatically, using CVC4 and Z3. Interestingly, the tools
complemented each other. Some goals that Z3 could not prove were
easily discharged by CVC4, and vice-versa. Only six goals were completed manually in Coq.

There are three instances of a loop invariant with an existential quantifier. The automated provers were unable to prove that the invariant holds initially. The proof is trivially completed with two lines in Coq - first, specifying an appropriate value, and then using the \texttt{auto} tactic.

There is one complex inequality -- the property of the expected value
calculation mentioned earlier. This requires a quite involved Coq
proof of about 120 lines.

Finally, there was a function with a fairly complex loop invariant that the automated provers could not quite verify. There are two goals, for the two possible outcomes of a branch within the loop. Each goal requires a fairly simple Coq proof, of about 15 lines. We present one of these proofs as an example in Appendix \ref{proof}.
      
The result is some confidence that the code is free from bugs. Unfortunately, this guarantee rests on the assumption that Krakatoa has no soundness errors. As discussed below, this assumption is not as safe as one would like.

Additionally, Krakatoa models floating-point variables as mathematical real numbers, so there are no guarantees about floating-point overflow or rounding. This model is still superior to anything provided by other Java tools.

Finally, through a special annotation, we instructed Krakatoa to ignore integer overflow. This made the proving process much simpler, but is another source of potential unsoundness. In particular, the model \emph{would} overflow if it is allowed to run for enough cycles, but it generally does not run that long.

\section{Potential for improvement}

We believe that wider adoption of formal methods is desirable, because it would improve the quality of software. For this to happen, the tools need to become easier to use; both so that existing users could apply formal methods more widely, and to make them accessible to a greater population. Based on our experience with Krakatoa, we have identified several areas in which we think more work would be particularly fruitful. We note that our list does \emph{not} include improving the power of automated SMT solvers, which would initially seem the most obvious approach to making machine-assisted proof work easier.

\subsection{Maintenance}

Many of these tools surveyed appear to be academic projects that were created and then abandoned. Some, like LOOP and JACK, are no longer accessible. Others do not support recent versions of Java. Installing old versions of Java is not only an extra burden during setup, it is also a security risk. Most tools were unexpectedly difficult to set up. Many of them had undocumented dependencies or broken install packages. They often produced unhelpful error messages when their dependencies were not satisfied. Finally, documentation was either lacking, incomplete, or severely outdated. We could not complete our proof in Krakatoa until we looked at the developers' test cases, which showcased several undocumented annotations.

If a hypothetical researcher wishes to create a tool that people will actually use, we recommend that they plan ahead to make sure they can document its features, and provide some rudimentary maintenance in the future. This does not mean supporting every new Java construct as it appears, but it would be an enormous improvement if the tools could at least parse more recent Java \texttt{*.class} files and provide precise error messages about unsupported features of Java.

\subsection{Bugs}

Krakatoa was the only tool we used for long enough to notice actual bugs in the program. Besides a slew of Java features that the tool did not support and which elicited obscure error messages, there were several of what we would call \emph{completeness errors}, where Krakatoa did not provide enough assumptions to prove true statements. An example is Krakatoa's treatment of class invariants. For each class invariant, Krakatoa generates pre- and post-conditions for each of the class's functions. It also generates appropriate pre-conditions as assumptions to functions that take elements of the class as parameters. However, Krakatoa does not assume that elements of an array satisfy the invariant. Thus, we had to manually add and prove the assertion that all elements of an array were valid at all times.

Completeness errors are not as bad as soundness errors, but their presence is a bad sign. Completeness errors stand out when we cannot complete a proof. However, we would not notice if one of Krakatoa's premises was unsound -- a proof of something we believe to be true would simply succeed where it should not. Thus, a formal verification tool with completeness errors gives far less confidence than it should in the soundness of the proof it produces.

\subsection{Floating point support}

IEEE floating point arithmetic is complicated, and it is unsurprising that most of the tools do not support it. However, most scientific computing uses floating point arithmetic. Hence, in the context of verifying a computational model, floating point support is essential. Modelling floating point variables as real numbers is not ideal, but is better than nothing. We see this shortcut as being akin to ignoring integer overflow -- it may be an acceptable approximation, as long as we are aware of its limitations.

Much worse than a lack of floating point support is \emph{fake} support. While TACO produced error messages informing us that it did not support floating point, Sireum/Kiasan and KeY claimed that they did. However, Sireum/Kiasan proved the false statement that \texttt{0.0 == 1.0}, and KeY was unable to prove that \texttt{0.0 == 0.0}. While $x=x$ is not a theorem for IEEE floating points ($\text{NaN} \neq \text{NaN}$), a special case like that should be provable if their behaviour is adequately modelled. A tool that cannot prove such a triviality is no better than one that does not support floating point arithmetic at all.

\subsection{Error messages}

Even before reaching the proving stage, many of the tools surveyed
failed to parse Java source or \texttt{*.class} files with unhelpful
error messages. \emph{Syntax error on line X} is bad (especially when
the correct syntax is undocumented), but \emph{assertion failure in
  parser} is even worse. Widely used compilers have become adept at
explaining their reasons for failure, and formal verification tools
should also aim to do so.

\subsection{Comprehensible proof goals}

By the end of the process, we discovered that automated SMT solvers
could prove almost all of the statements of interest about our
program. However, this was only after a long time spent adjusting the
JML annotations. At the beginning of the process, it seemed that automated tools were hopelessly inadequate, and that better solvers or interactive provers were necessary. This can create the illusion that more progress in automated proving is needed to make formal methods more accessible. But it is our opinion that two completely different factors are crucial. The first is reasoning about missing assumptions, which we discuss in the next section. The other is comprehensible proof goals.

By comprehensible goals we mean that formal verification tools today do little to help the user understand the details of the formal logical statement they are trying to prove. This statement is the result of a complex transformation process peculiar to each tool, and may be very different from what the user expects based on the source code they are working with. Often, understanding the proof goal is enough to see which annotations need to be changed.

The most incomprehensible goals come from the fully automated tools like OpenJML. These tools opaquely convert annotations into proof goals, send them to SMT solvers, and fail. Because the user cannot see the unprovable goal, they do not know why the proof failed. Appendix \ref{loop} gives an example of a mistake that may trip up novices, and which tools should be able to prevent.

Some tools do display their proof goals. The interactive tools, KeY and Krakatoa, provide a full proof script that the user can examine. This is an improvement, as it often becomes clearer why a proof is failing once one has investigated the obligation closely, and perhaps even attempted an interactive proof. However, it is not ideal, because the language in which the proof goal is expressed is very far removed from the Java code it is based on, and important details are often obscured by copious boilerplate.

We feel that making an effort to explain proof goals to the user in terms of the code they are working with would make the tools enormously more accessible. This means expressing the statements being proved not in terms of complex mangled names for constructs from the tool's memory model, but in terms of Java objects, and lines of Java code. In Appendix \ref{assumption} we include an example of what this would look like in our ideal world. 

\subsection{Inferring missing assumptions}

As mentioned earlier, most of our program was successfully verified with automated tools, but only after a long process of editing the annotations. Most of these edits involved adding assumptions that were missing from the proof goal, such as function pre-conditions and loop invariants. We added them after looking closely at the generated proof goal, or attempting an interactive proof and discovering the missing hypothesis. This bears repeating: the vast majority of the statements that the SMT solvers could not prove were in fact false. More precisely, they were true when expressed as JML assertions, but the proof obligations generated by the tool were false, as they lacked information not captured by the annotations. The main barrier to completing the proof automatically was not the power of the provers, but inadequate annotations. We present an example in appendix \ref{assumption}.

With SMT solvers doing the proving, most of our effort was spent on inferring which assumptions were missing from our JML annotations. Thus, we believe the greatest potential for making formal verification more user friendly is in finding ways to automatically answer the question \emph{which extra assumptions would make this goal true?} This is the problem referred to as \emph{abduction} \cite{abduction1}, and we suffer no illusions as to its difficulty. Abductive reasoning has been studied in logic \cite{abductionlogic}, philosophy \cite{abductionphil} and artificial intelligence \cite{abductionai} as a means of inferring the \emph{cause} of an event. Some such capability added to formal verification tools might allow them to make suggestions for extra annotations. Appendix \ref{assumption} contains a concrete example.

We are aware of one static analysis tool, Facebook's INFER \cite{infer}, that uses a form of abduction. The INFER tool checks several common error conditions such as null pointer dereferences, and does not require any annotations, because it infers pre- and post-conditions automatically. We have not investigated INFER in this paper because it does not verify user-specified post-conditions or assertions.

\section{Conclusion}

Tools for formal verification of Java programs are in generally poor shape, and do not support recent versions of Java. Nonetheless, they do exist, and are usable enough to verify a program. SMT solvers are powerful enough to prove most true statements of interest; the most difficult aspect of verification was, in our experience, finding the right annotations to use to ensure the provers had enough assumptions to work with. We believe wider adoption of formal methods would be beneficial, and making the tools more user-friendly will bring about this goal. Existing tools could use better documentation, error messages, and fewer bugs. We also think a large usability gain is waiting to be realised by tools that can clearly relate the proof goals they generate to the source code the user is working with, and that can suggest missing assumptions to the user that would make a proof succeed.

%\balancecolumns

\bibliographystyle{abbrv}
\bibliography{references}

\newpage
\appendix
\section{Proof Goal Example} \label{loop}

When proof goals are opaque, there is no way to know if an SMT solver's failure was because the statement is false, because the statement is very very difficult, or because the assumptions need a trivial strengthening. For example, the SMT tools could not automatically prove that the code below would not cause an \texttt{ArrayIndexOutOfBoundsException}:

\begin{verbatim}
//@ invariant (\forall int j;
                0 <= j < i ==> array[j] == 0);
for(int i = 0; i < array.length; i++)
    array[i] = 0;
	
\end{verbatim}

The solution is to add \texttt{0 <= i <= array.length} to the
invariant. Those experienced in formal methods will see this
immediately, but novices may be baffled by
the tools' inability to prove simple statements.

If the tool showed the user the proof goal, along with its assumptions, they should be able to spot the error quickly.

\section{Missing Assumption Example} \label{assumption}

A prover that could report \emph{I could complete this proof, if only I knew that P holds}, would represent an enormous usability improvement. We present here a simple example of the obscurity of the proof goals generated by Krakatoa, and of the kind of suggestion we would like to see from an ideal prover with abductive reasoning capability.

The following code:

\begin{verbatim}
class Example {
  int[] field;
  
  void doNothing() {
    return;
  }

  static void func(Example object) {
    object.field = new int[10];
    object.doNothing();
    //@ assert object.field.length == 10;
  }
}
\end{verbatim}

Would produce a long and complicated proof goal filled with boilerplate, which no SMT solver will prove. But somewhere in there would be something like this:

\begin{verbatim}
Axiom:

(Jessie_memory_model.offset_max
  usObject_alloc_table
    (Jessie_memory_model.select
      usField object)) + 1 = 10

Proof Goal:

(Jessie_memory_model.offset_max
  usObject_alloc_table1
    (Jessie_memory_model.select
      usField object)) + 1 = 10
\end{verbatim}

It takes experience with Krakatoa to understand this. The small change in \texttt{usObject\_alloc\_table} is the culprit. The statement expresses the proposition 
\begin{verbatim}
object.field.length == 10
\end{verbatim}
More specifically, the goal is to prove that the array still has the same length as it did earlier. Because the \texttt{doNothing()} function is not annotated, Krakatoa will not make any assumptions about the value of fields after it is called. Thus, it is impossible to prove anything about their value. To fix this, we should annotate \texttt{doNothing()} with details about which fields it changes, and which fields remain the same.

Ideally, when this proof fails, the prover should report that it needs the extra assumption
\begin{verbatim}
(Jessie_memory_model.offset_max
  usObject_alloc_table
    (Jessie_memory_model.select
      usField object))
= (Jessie_memory_model.offset_max
    usObject_alloc_table1
      (Jessie_memory_model.select
        usField object)).
\end{verbatim}
Even better, the tool should be able to relate this to actual Java code, and report something like \emph{I cannot prove that \emph{\texttt{object.field.length}} at line 9 of the file is equal to \emph{\texttt{object.field.length}} at line 11}.

\section{Coq Proof Example} \label{proof}
This is an example of the short Coq proof for one of the loop invariants. This is the loop in question, edited for clarity:

\begin{verbatim}
int best = 0;
/*@
loop_invariant
(\forall int j; 0 <= j < h ==>
  expectedValue(j) <= expectedValue(best))
@*/
for (int h = 1; h < ATTACK_TYPES; h++) {
  if (expectedValue(h) > expectedValue(best))
    best = h;
}
\end{verbatim}

The invariant produces three obligations. One to prove that it holds initially, and one to prove that it holds after executing the body where the \texttt{if} block is not entered, and one where the block \emph{is} entered. We present the third of these, where the block is entered.

The proof script itself is the file ending in\\
\texttt{...highestExpectedValue\_ensures\_default\_5.v}\\
in the git repository \url{https://github.com/dmitry-brizhinev/formal-verification}. It contains copious amounts of boilerplate. We present here a version heavily edited for conciseness and clarity:

\begin{verbatim}
Variable j:Z, h':Z, h:Z, best_old:Z, best_new:Z

Premise p1: forall (j':Z), (0 <= j' < h')
     expectedValue j' <= expectedValue best_old
Premise p2: expectedValue best_old < expectedValue h'
Premise p3: best_new = h'
Premise p4: h = h' + 1
Premise p5: 0 <= j < h

Goal: expectedValue j <= expectedValue best_new

Proof:
rewrite p3.
cut (0 <= j < h' \/ j = h').
intro.
destruct H.
cut (expectedValue j <= expectedValue best_old).
intro.
apply Rle_trans with (r2 := expectedValue best_old).
trivial.
auto with *.
auto.
rewrite H.
auto with *.
omega.
Qed.
\end{verbatim}

The key elements of the proof are:
\begin{enumerate}
\item The first cut, splitting the proof into the cases \texttt{j < h'} and \texttt{j = h'}.
\item Destruct, which considers the two cases separately.
\item The second cut, which instantiates premise \texttt{p1} with \texttt{j' = j}.
\item The use of the transitivity of less-than, which connects premises \texttt{p1} and \texttt{p2}
\item Omega, which automatically solves some inequalities.
\end{enumerate}

\end{document}